\documentclass[preprint,proceedings]{rmaa}
\usepackage{paralist}

\usepackage{psfrag,color}

\SetYear{2002}
\SetConfTitle{Winds, Bubbles, and Explosions}

\title{An Inside Out View of Bubbles} 

\author{
  You-Hua Chu,\altaffilmark{1} 
  Robert A.\ Gruendl,\altaffilmark{1}
  and Mart\'{\i}n A.\ Guerrero\altaffilmark{1}}

\altaffiltext{1}{Astronomy Department, University of Illinois, 
  1002 W.\ Green Street, Urbana, IL 61801, USA}

\shortauthor{Chu, Gruendl, \& Guerrero}
\shorttitle{An Inside out View of Bubbles}

\fulladdresses{
\item You-Hua Chu, Robert A. Gruendl, and Mart\'{\i}n A. Guerrero:
  Astronomy Department, University of Illinois, 1002 W.\ Green Street,
  Urbana, IL 61801, USA (chu@astro.uiuc.edu, Gruendl@astro.uiuc.edu,
  mar@astro.uiuc.edu).}

\listofauthors{
Y.-H. Chu, R. A. Gruendl, M. A. Guerrero}
\indexauthor{Chu, Y.-H.}
\indexauthor{Gruendl, R. A.}
\indexauthor{Guerrero, M. A.}
\abstract{
Fast stellar winds can sweep up ambient media and form bubbles.
The evolution of a bubble is largely controlled by the content 
and physical conditions of the shocked fast wind in its interior. 
This hot gas was not clearly observed until the recent advent of
{\it Chandra} and {\it XMM-Newton} X-ray observatories.
To date, diffuse X-ray emission has been unambiguously detected
from two circumstellar bubbles blown by WR stars, four planetary 
nebulae, and two superbubbles blown by young clusters.
Model fits to the X-ray spectra show that the circumstellar
bubbles are dominated by hot gas with low temperatures ($\le 3\times
10^6$ K), while the interstellar bubbles contain significant 
fractions of hotter gas ($\ge 5\times10^6$ K).
In all cases, large discrepancies in the X-ray luminosity are 
found between observations and conventional models of bubbles.
Future theoretical models of bubbles need to re-examine the 
validity of heat conduction and take into account realistic 
microscopic processes such as mass loading from dense clumps/knots 
and turbulent mixing.  {\it Chandra} ACIS-S observation of NGC\,6888 
will shed light on these astrophysical processes.}

\addkeyword{H\,II regions}
\addkeyword{ISM: bubbles}
\addkeyword{H~II regions}
\addkeyword{Planetary nebulae}
\addkeyword{Stars: Mass loss}
\addkeyword{Stars: Wolf-Rayet stars}
\addkeyword{X-rays}

\begin{document}
% Typeset article header
\maketitle

\section{Introduction: A Brief History of Bubble Studies}

In 1965, Johnson \& Hogg reported three shell nebulae, 
NGC\,2359, NGC\,6888, and the nebula around HD\,50896,
and suggested that these nebulae were formed by  
interactions between the mass ejected by the central 
Wolf-Rayet (WR) stars and the ambient interstellar gas.
In modern terminology, these shell nebulae are called
``bubbles", the mass ejected by a WR star is ``fast 
stellar wind", the shell nebula around HD\,50896 has been 
cataloged as ``S\,308", and the ambient medium is 
designated ``circumstellar medium", i.e., ejected stellar 
material.

The earliest theoretical treatments of interactions between 
fast stellar winds and the interstellar medium (ISM) were
motivated by the central cavities and high-velocity 
motions observed in \ion{H}{2} regions (Mathews 1966; Pikel'ner 
1968; Pikel'ner \& Schcheglov 1969; Dyson \& de Vries 1972). 
Models of wind-ISM interaction were constructed specifically 
for shell nebulae around WR stars by Avedisova (1972),
and a numerical calculation of wind-ISM interaction was 
provided by Falle (1975).

In 1974, Jenkins \& Meloy reported {\it Copernicus} 
satellite observations of 32 early-type stars, which 
revealed ubiquitous shallow interstellar \ion{O}{6} 
absorption.  To explain this \ion{O}{6} absorption,
Castor, McCray, \& Weaver (1975) modeled wind-ISM 
interaction and coined the term ``interstellar bubble".
Weaver et al.\ (1977) elaborated on this model and
gained popularity because of their detailed description
of the physical conditions and structure of a bubble 
and their specific characterization of X-ray emission and
\ion{O}{6} column density expected from a bubble interior.

High-quality, high-resolution X-ray and far UV 
observations were not possible until {\it Chandra X-ray
Observatory}, {\it XMM-Newton X-ray Observatory}, and 
{\it Far Ultraviolet Spectroscopic Explorer (FUSE)} 
satellites were launched.
It is finally possible to observe the hot gas in bubble
interiors and critically examine the bubble models.
The analysis of {\it FUSE} observations is complex, and
it is pre-mature to conclude on the \ion{O}{6} absorption
from bubbles.  Therefore, in this paper we will concentrate
on only X-ray observations of bubbles and compare them 
to model predictions.

\begin{table*}[!t]\centering
  \newcommand{\DS}{\hspace{6\tabcolsep}} %% Expanded Space between
  %% some cols
  \setlength{\tabnotewidth}{0.9\textwidth}
  \setlength{\tabcolsep}{1.33\tabcolsep}
  \tablecols{5}
  % Stretch the space between table columns 
  \setlength{\tabcolsep}{1.5\tabcolsep}
  \caption{X-ray Observing Facilities}
  \begin{tabular}{clccc}
    \toprule
    X-ray    & ~Mission  & Imaging & Angular    &  Energy \\
    Observatory  & Duration & Spectrometer\tabnotemark{a}  & Resolution & Range (keV)\\
    \midrule
   {\it Einstein}  & 1978-1981  &  IPC          &  120$''$ & 0.2-3.5 \\
   {\it ROSAT}     & 1990-1998  &  PSPC         &  30$''$  & 0.1-2.4 \\
   {\it ASCA}      & 1993-2001  &  SIS          &  150$''$ & 0.4-10  \\
   {\it Chandra}   & 1999-pres. &  ACIS         &  1$''$   & 0.1-10  \\
   {\it XMM-Newton}& 1999-pres. &  EPIC         &  10$''$  & 0.1-15  \\
    \bottomrule
    \tabnotetext{a}{IPC: Imaging Proportional Counter; PSPC: Position Sensitive
        Proportional Counter; SIS: Solid-State Imaging Spectrometer; 
        ACIS: Advanced Camera for Imaging and Spectroscopy; EPIC: European 
        Photon Imaging Camera}
  \end{tabular}
\end{table*}

\section{Bubble Models}

Most of the early models of bubbles assume that the stellar
wind interacts with the ambient medium through momentum 
transfer and find that the shell expansion follows 
$r \propto t^{1/2}$, where $r$ is the shell radius and $t$ 
is the dynamic age (Mathews 1966; Pikel'ner 1968; Avedisova
1972; Steigman, Strittmatter, \& Williams 1975).
Their assumption of momentum conservation may be convenient for
1-D calculations in which stellar wind does not get deflected. 
However, in the 3-D space, momentum is a vector and the momentum 
flux of a stellar wind is null (otherwise the star itself would 
experience a rocket effect and gain momentum); it is unphysical 
to assume that the wind momentum in every specific direction is 
conserved.

\begin{figure}[!b]
  \includegraphics[width=\columnwidth]{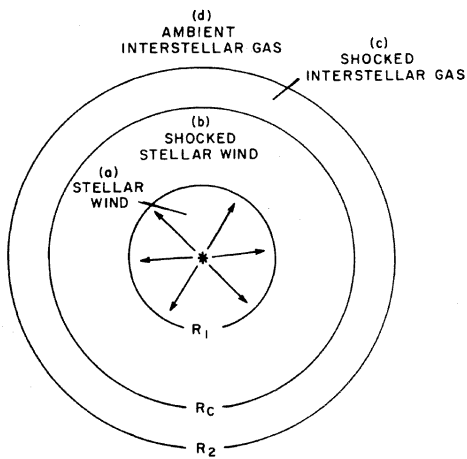}
  \caption{Schematic structure of a pressure-driven interstellar
   bubble, from Weaver et al.\ (1977).}
  \label{fig:bub_cartoon}
\end{figure}

Dyson \& de Vries (1972) were the first to suggest that the 
expansion of a bubble is driven by the thermal pressure of 
the shocked wind in the bubble interior, and follows the 
$r \propto t^{3/5}$ law.  
This formed the basis of the interstellar bubble model by
Castor, McCray, \& Weaver (1975), who implemented heat 
conduction at the interface between the hot interior and 
the cool swept-up shell.

\begin{figure}[!t]
  \includegraphics[width=\columnwidth]{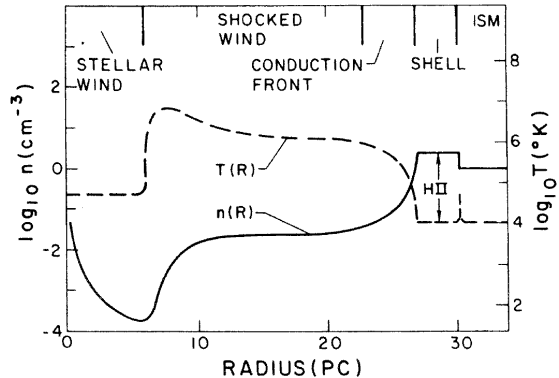}
  \caption{Temperature and density profiles of  a 
   pressure-driven interstellar bubble, from Weaver et al.~(1977).}
  \label{fig:bub_DT}
\end{figure}

The basic structure of an interstellar bubble, illustrated
in Figure 1, can be divided into four zones: (a) freely
expanding stellar wind, (b) shocked stellar wind, (c) swept-up
ISM, and (d) ambient ISM.  The temperature and density structure 
of a bubble is illustrated in Figure 2.   
Two shocks are present in a bubble, an adiabatic shock
at $R_1$, between zones (a) and (b), and an isothermal 
shock at $R_2$, between zones (c) and (d).  The hot,
shocked stellar wind and the swept-up ISM are separated by a 
contact discontinuity at $R_{\rm c}$.
In the shocked stellar wind layer, the temperature structure is
modified by heat conduction, and the density at the outer edge
is raised by mass evaporation across $R_{\rm c}$ from the dense 
swept-up interstellar shell.
At the conduction front, the temperatures are high enough to
produce collisionally ionized O$^{+5}$, which may be responsible
for the interstellar \ion{O}{6} absorption detected in {\it
Copernicus} observations of early-type stars.

For a pressure-driven bubble in a homogeneous ISM, the
X-ray emissivity can be integrated over the volume with 
X-ray-emitting temperatures to determine the expected X-ray
luminosity in soft energy band.
As given by Chu et al.\ (1995), the expected X-ray luminosity 
in the 0.1--2.4 keV band can be expressed in wind parameters as 
$L_{\rm X} \approx 
(2 \times 10^{35}$ ergs s$^{-1} ) \xi L_{37}^{33/35} n_0^{17/35} t_6^{19/35}$,
where $\xi$ is the metallicity relative to the solar value, $L_{37}$ 
is the mechanical luminosity of the stellar wind in units of 10$^{37}$ 
ergs s$^{-1}$, $n_0$ is the ambient density in cm$^{-3}$, and $t_6$ is the 
dynamic age in units of 10$^6$ yr.
The X-ray luminosity can also be expressed in bubble parameters as
$L_{\rm X} \approx (1.6 \times 10^{28}$ ergs s$^{-1} ) \xi  n_0^{10/7} 
R_{\rm pc}^{17/7} V_5^{16/7}$, where $R_{\rm pc}$ is the shell
radiua in units of pc, and $V_5 = 0.59 R_{\rm pc} / t_6$ is the
shell expansion velocity in units of km s$^{-1}$.

Shell nebulae around WR stars are circumstellar bubbles consisting of 
stellar material ejected by the progenitors at a red supergiant (RSG) 
phase or a luminous blue variable (LBV) phase; the circumstellar 
medium is far from homogeneous.
The formation and evolution of WR bubbles have been hydrodynamically
simulated by Garc\'{\i}a-Segura et al.\ (1996a,b) for WR stars that 
have evolved through LBV and RSG phases, respectively.  
These models follow the same basic principles as Weaver et al., but have
encorporated realistic stellar mass loss history in their calculations.
While these models are successful in reproducing the nebular morphologies,
they find that the stellar wind luminosity expected from the observed
bubble dynamics is often more than an order of magnitude lower than
that derived from observations of the wind directly.  

The clumpy morphology of WR bubbles implies that the stellar wind 
may be interacting with small fragments of nebular material, thus
mass loading may be an important process that modifies the physical
conditions of a bubble interior.
Adiabatic bubbles with conductive evaporation and hydrodynamic
ablation have been modeled by Pittard et al.\ (2001a, b).
They find the X-ray emissivity and temperature profiles of a
bubble interior can rise or fall toward the outer edge, depending
on the ratio of wind mass to injected nebular mass.

Wind-blown bubbles are usually associated with massive stars;
however, the formation of planetary nebulae (PNe) around
low- and intermediate-mass stars is almost identical to that 
of a WR bubble (e.g., Kwok 1983; Chu 1993).
Therefore, we will include PNe in our discussions in this paper.

\section{X-ray Observations of Bubbles}

X-ray observations of wind-blown bubbles and PNe have been made 
with a number of satellites, of which the duration of the mission, 
instrument used, angular resolution, and photon energy range are
summarized in Table 1.  {\it Einstein} IPC observations detected
diffuse X-ray emission from only one wind-blown bubble, NGC\,6888
(Bochkarev 1988).  Many detections of diffuse X-ray emission from 
interstellar bubbles in HII regions based on {\it Einstein} 
observations, e.g., the Orion Nebula (Ku \& Chanan 1979), the 
Rosette Nebula (Leahy 1985), S155 (Fabian \& Stewart 1983), were 
later shown by {\it ROSAT} observations to be spurious, as the 
X-ray emission was resolved into stellar point sources (Chu 1994).

In Table 2, we summarize X-ray observations of bubbles and
PNe made with ``modern" X-ray satellites within the last decade.
Objects with diffuse X-ray emission detected are listed at the
top without parentheses, and the non-detections are listed at
the bottom in parentheses. 

\begin{table}[!b]\centering
  \setlength{\tabnotewidth}{\columnwidth}
  \tablecols{3}
  % Stretch the space between table columns 
  \setlength{\tabcolsep}{1\tabcolsep}
  \caption{X-ray Observations of Bubbles and Planetary Nebulae\,\tabnotemark{a,b}}
  \begin{tabular}{lll}
    \toprule
    {\it ROSAT} &  {\it Chandra}   & {\it XMM-Newton}\\
    \midrule
\multicolumn{3}{c}{\it ---~~~Bubbles~~~---} \\
     NGC\,6888         & Rosette Nebula    & S\,308    \\
     S\,308              & Omega Nebula      &         \\
     Omega Nebula      &                   &           \\
     (NGC\,2359)       &                   &           \\
     (NGC\,3199)       &                   &           \\
     (NGC\,6164-5)     &                   &           \\
     (NGC\,7635)       &                   &           \\
\midrule
\multicolumn{3}{c}{\it ---~~~Planetary Nebulae~~~---} \\
%\multicolumn{3}{c}{-- -- -- -- -- -- -- -- -- -- -- -- -- -- -- 
%-- -- -- -- -- -- -- -- -- --} \\
     BD+30$^\circ$3639 & BD+30$^\circ$3639 & NGC\,7009 \\
     NGC\,6543         & NGC\,6543         &           \\
     Abell 30          & NGC\,7027         &           \\
%\multicolumn{3}{c}{-- -- -- -- -- -- -- -- -- -- -- -- -- -- -- 
%-- -- -- -- -- -- -- -- -- --} \\
     (60+ PNe)         & (NGC\,7293)       &           \\
                       & (He\,2-90)        &           \\
                       & (M\,1-16)         &           \\
    \bottomrule
\tabnotetext{a}{Diffuse X-ray emission is detected only from objects 
    listed at the top without parentheses.}
\tabnotetext{b}{References for objects with diffuse X-ray emission --
    Abell\,30: Chu, Chang, \& Conway (1997), Chu, \& Ho (1995); 
    BD+30\arcdeg 3639: Kastner et al.\ (2000); NGC\,6543: Chu et al.\ (2001); 
    NGC\,6888: Wrigge, Wendker, \& Wisotzki (1994); NGC\,7009: Guerrero, 
    Gruendl, \& Chu (2002); NGC\,7027: Kastner, Vrtilek, \& Soker (2001); 
    Omega Nebula: Townsley et al.\ (2002), Dunne et al.\ (2002, in prep.); 
    Rosette Nebula: Townsley et al.\ (2001); S\,308: Wrigge (1999), Chu et al.\
    (2002, in prep.).}
   \end{tabular}
\end{table}

\subsection{Diffuse X-ray Emission from WR Bubbles}

Diffuse X-ray emission has been detected from two WR bubbles,  
NGC\,6888 and S\,308.  Figures 3 and 4 show optical [O III]
and {\it ROSAT} PSPC images of these two bubbles (Wrigge,
Wendker, \& Wisotzki 1994; Wrigge 1999).  
Despite the limited angular resolution, the PSPC X-ray image 
of NGC\,6888 shows clear limb-brightening.
The X-ray morphology of S\,308 cannot be easily visualized
because the shell rim is occulted by the circular ring of 
the PSPC's window support structure.
Analyses of the spectra extracted from these PSPC observations 
show that the hot gas in the interiors of these two WR bubbles
is dominated by gas at $\sim$1.5$\times$10$^6$ K.  
Wrigge (1999) also suggested a high-temperature, 2.8$\times$10$^7$ K, 
component in S\,308; however, this component may be contributed
by the numerous unresolved point sources.
{\it ASCA} observations of NGC\,6888 detected an additional 
plasma component at $\sim$8$\times$10$^6$ K (Wrigge et al.\ 1998).

These observations can be greatly improved by {\it Chandra} and
{\it XMM-Newton}.  We have been awarded a 100 ks
{\it Chandra} observation of NGC\,6888 in Cycle 4 that will be
made in 2003.  We have obtained 40 ks {\it XMM-Newton} observations
of S\,308, with the field of view covering the northwest quadrant
of the bubble (see Figure 5).
Although these {\it XMM} observations were affected by  
high background during 2/3 of the total exposure time, they
show clearly a limb-brightened distribution of the X-ray emission
from the interior of S\,308.
Most interestingly, a spatial gap is present between the outer
edge of diffuse X-ray emission and the outer edge of [O III] 
emission, and this gap might be the conduction front
that has been long sought after!
The X-ray spectrum of S\,308 (see Figure 6) is extremely soft,
indicating a plasma temperature of only $\sim$1$\times$10$^6$ K.

\begin{figure}[!t]
  \includegraphics[width=\columnwidth]{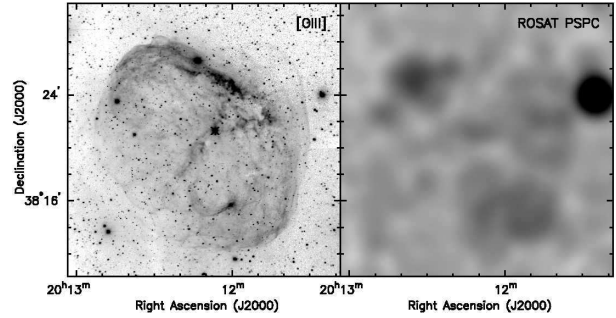}
  \caption{Optical [O III] $\lambda$5007 image (left) and
   {\it ROSAT} PSPC image (right) of NGC\,6888.}
\end{figure}

\begin{figure}[!t]
  \includegraphics[width=\columnwidth]{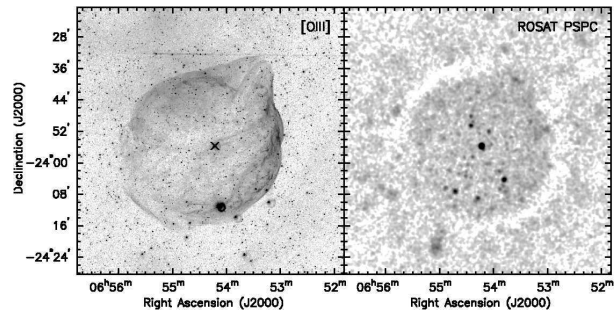}
  \caption{Optical [O III] $\lambda$5007 image (left) and
   {\it ROSAT} PSPC image (right) of S\,308.}
\end{figure}

\begin{figure}[!t]
  \includegraphics[width=\columnwidth]{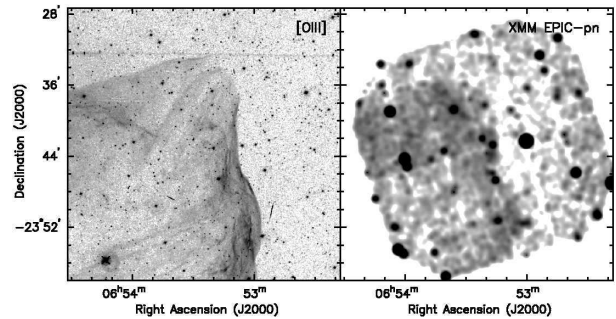}
  \caption{Optical [O III] $\lambda$5007 image (left) and
   {\it XMM-Newton} EPIC image (right) of the northwest 
   quadrant of S\,308.}
\end{figure}

\subsection{Diffuse X-ray Emission from Planetary Nebulae}

\begin{figure*}[!t]
\begin{center}
  \includegraphics[width=0.7\textwidth]{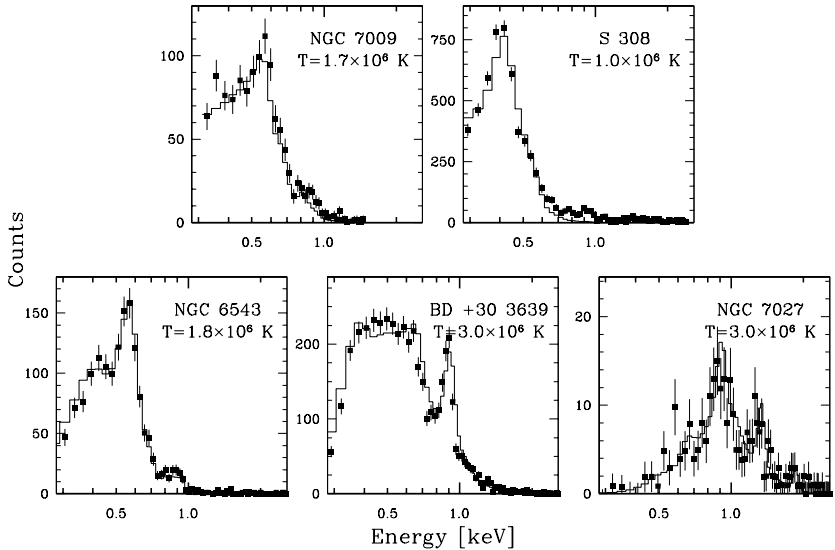}
  \hspace*{\columnsep}%
  \caption{{\it XMM-Newton} EPIC-pn and $Chandra$ ACIS-S spectra of 
   four PNe and one WR bubble.  }
\end{center}
\end{figure*}

More than 60 PNe have been observed by {\it ROSAT}, but only
three nebulae show marginally extended X-ray emission
(Guerrero, Chu, \& Gruendl 2000). 
{\it Chandra} observations unambiguously resolved the diffuse 
X-ray emission from three PNe, and  {\it XMM-Newton}
observations resolved the diffuse X-rays from NGC\,7009.
See Table 2 for the names and references of these PNe.
Figure 7 shows the most well-resolved PN, NGC\,6543 
(the Cat's Eye Nebula), where a limb-brightened morphology is 
clearly seen.  

The X-ray spectra of the diffuse emission from four PNe are 
displayed in Figure 6; all show thermal plasma emission.
The best-fits to these spectra indicate plasma temperatures of 
2--3$\times$10$^6$ K and densities of $\sim$100 cm$^{-3}$, from
which we conclude that the hot gas in PN interiors is
over-pressurized and drives the expansion of the optical nebular
shell.

\begin{figure}[!h]
  \includegraphics[width=\columnwidth]{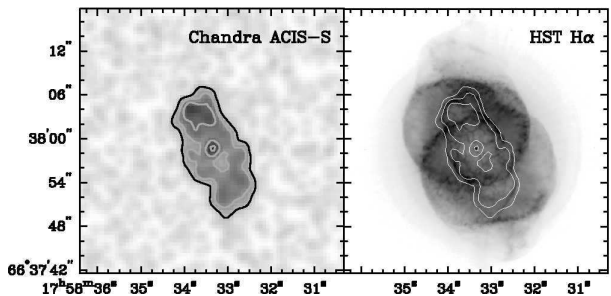}
  \caption{{\it Chandra} ACIS image (left) and {\it HST} WFPC2
    H$\alpha$ image overlaid by X-ray contours (right) of the PN
    NGC\,6543, the Cat's Eye Nebula.}
\end{figure}

\subsection{Diffuse X-ray Emission from Superbubbles}

Diffuse X-ray emission from quiescent superbubbles (without
recent supernova blasts) is detected for the first time
by {\it Chandra} -- the Rosette Nebula and the Omega
Nebula (Townsley et al.\ 2001, 2002).  
The diffuse X-ray emission from the Omega Nebula was 
in fact detected by {\it ROSAT}, but has never been reported.
Figure 8 shows the {\it ROSAT} PSPC image and an optical 
image of the Omega Nebula; the diffuse X-ray emission fills 
the entire interior of this superbubble.
As the ionizing cluster of the Omega Nebula is only 
$\sim$1$\times$10$^6$ yr old, the hot gas must be 
produced solely by fast stellar winds.

The diffuse X-ray emission from the two interstellar 
superbubbles, the Rosette Nebula and the Omega Nebula,
is qualitatively different from the diffuse emission
from circumstellar bubbles, i.e., WR bubbles and PNe.
First of all, the diffuse emission from the superbubbles 
does not show any limb-brightening.  Second, X-ray
spectra of the superbubbles show a high-temperature
($\sim$7$\times$10$^6$ K) component in addition to the 
dominant component at $\sim$2$\times$10$^6$ K.
Finally, the densities of the hot gas in superbubbles are 
much lower than those in the interiors of WR bubbles or PNe.

\begin{figure}[!h]
  \includegraphics[width=\columnwidth]{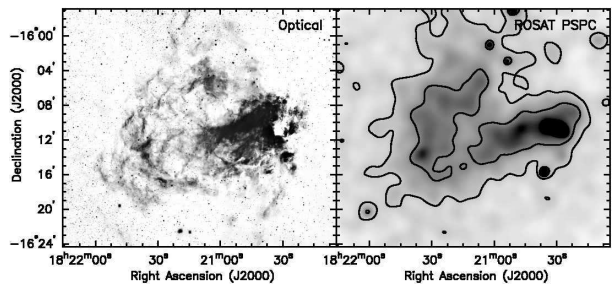}
  \caption{Digitized Sky Survey red-band image (left) and
   {\it ROSAT} PSPC image (right) of the Omega Nebula.}
\end{figure}

\section{Observations vs. Theoretical Expectations}

X-ray observations have detected diffuse emission from two
WR bubbles, four PNe, and two young superbubbles.  
The typical physical properties of the X-ray-emitting gas derived
from {\it Chandra} and {\it XMM-Newton} observations are listed
in Table 3.
The average X-ray spectrum of a bubble is weighted toward
the densest X-ray-emitting region.
As shown in Figure 2, the density and temperature profiles
in a bubble interior are anti-correlated, thus the brightest
emission is expected to originate from regions with the lowest
temperatures.
The observed low plasma temperatures and limb-brightened morphology
are fully consistent with this expectation.
The observed variations of plasma density among the three
types of bubbles are also qualitatively consistent with
the expectation from mass evaporation processes, as the plasma 
density should reflect the nebular density.

However, two outstanding discrepancies exist between
the observations and theoretical predictions when they
are compared {\it quantitatively}.
First, the observed X-ray luminosities are 10--100 times lower
than those predicted by pressure-driven bubble models using
observed bubble dynamics, such as shell size, expansion velocity,
nebular density, and stellar wind mechanical luminosity.
Second, in a pressure-driven bubble model, the hot gas near 
the conduction front is dominated by nebular mass evaporated 
across the contact discontinuity, but the abundances of
the X-ray-emitting gas in at least two PNe are consistent with 
those of the fast stellar winds instead of those of the nebular 
shells (Chu et al.\ 2001; Arnaud et al.\ 1996).
To resolve these problems, high-resolution observations 
of more bubbles are needed to 
study the detailed astrophysical processes, e.g., 
conduction evaporation, hydrodynamic ablation, and turbulent 
mixing.  Our upcoming 100 ks {\it Chandra} observation of 
NGC\,6888 may shed light on these processes.

\begin{table}[!t]\centering
  \setlength{\tabnotewidth}{\columnwidth}
  \tablecols{4}
  % Stretch the space between table columns 
  \setlength{\tabcolsep}{\tabcolsep}
  \caption{Physical Properties of Hot Gas in Bubble Interiors}
  \begin{tabular}{cccc}
    \toprule
    Bubble Type &  $T_{\rm e}$   & $N_{\rm e}$   &  $L_{\rm X}$ \\
       &  [10$^6$ K]    &  [cm$^{-3}$]  &   [ergs s$^{-1}$] \\
    \midrule
    Planetary Nebula   &  2--3   &  100    & 10$^{31} - 10^{32}$  \\
    WR Bubble          &  1, 8?  &   10    & 10$^{33} - 10^{34}$  \\
    Superbubble        &  2, 7   &   0.1   & 10$^{33} - 10^{34}$  \\
    \bottomrule
   \end{tabular}
\end{table}

\end{document}